\documentstyle[mprocl,epsfig]{article}

\def\be{\begin{equation}}
\def\ee{\end{equation}}
\def\bea{\begin{eqnarray}}
\def\eea{\end{eqnarray}}

\begin{document}
\begin{titlepage}
\begin{flushright}
CERN-TH/97-160 \\
July 1997\\
\end{flushright}
\vspace{2cm}
\centerline{\Large Two-Cutoff Lattice Regularization of Chiral Gauge Theories \footnote{Invited Talk at AIJIC 97 on Recent Developments in Non-perturbative Methods.}}
\vspace{10pt}
\centerline{\Large\bf }
\vspace{24pt}
\centerline{\large P. Hern\'andez}
\centerline{\it Theory Division, CERN}
\centerline{\it CH-1211 Geneva 23, Switzerland}

\vspace{4cm}
\begin{abstract}
I review our method to formulate chiral gauge theories 
on the lattice based on a two-cutoff lattice regularization and
discuss recent numerical results in a chiral $U(1)$ gauge theory in 2D.
\end{abstract}
\end{titlepage} 

\section{Introduction}

The well-known theorem by Nielsen and Ninomiya (NN) sets very strong 
constraints on the possibility of constructing a non-perturbative regulator
with an exact chiral symmetry.  On the lattice, the difficulty is related
to the presence of doubler modes, which render any symmetric discretization of
a chiral theory effectively vector-like. The doublers can be given masses of 
the order
of the cutoff, and thus be decoupled in the continuum limit, but only at the 
expense of explicitly breaking the chiral
symmetry. In the case of a chiral gauge theory, this implies the 
 explicit breaking of the gauge symmetry and it is
highly non-trivial to ensure that, in the continuum limit,  
the gauge symmetry is restored, without changing the particle
content of the model. The Roma approach \cite{roma}  
addresses this issue and provides a solution based
on a non-perturbative tuning of all dimension-four or less 
counterterms, necessary to ensure that BRST identities, which are broken 
at the regulator level, are satisfied in the continuum limit. Although the 
validity of this method is expected to hold
to all orders in perturbation theory, its practicability in real
simulations is questionable. 

I will review here an alternative method that we recently proposed \cite{us1,us2} to deal with chirality on the lattice. Our formulation of chiral gauge
theories involves two essential ingredients.

{\it Small breaking of gauge invariance at the cutoff level}. Even though the gauge symmetry
has to be broken at the cutoff level, according to NN, it is
still possible to constrain this breaking to be arbitrarily small. This is 
achieved by using a two-cutoff (TC) lattice formulation \cite{us1}, 
in which the 
fermion momenta are cut off at a scale much larger than the boson momenta.
More explicitly, the fermions 
live on a lattice of spacing $f$ and are 
coupled to gauge link variables that are constructed by an appropriate 
smooth and gauge-invariant interpolation  of gauge configurations 
\cite{luscher1}\cite{us2}
that are generated on a coarser lattice of spacing $b$. As long as the 
interpolation is smooth, the Fourier modes of the interpolated field
are effectively cut off at the scale $1/b \ll 1/f$. Doublers are decoupled 
by introducing a naive Wilson term as in the 
Roma approach \cite{roma}. However, the 
chirally breaking effects due to the Wilson term are relevant only at 
scales of the order of the fermion cutoff (the Wilson term is a higher-dimensional operator suppressed by the fermion cutoff), and
 gauge boson momenta 
are not large 
enough to prove these interactions, in contrast with one-cutoff (OC) formulations. As a result, due to the separation of 
the cutoff scales, it can be shown that no fine-tuning is necessary  
to recover an approximate chiral (global or gauge) symmetry.
The parameter $f/b$, which can be chosen to be arbitrarily small in principle,  
controls the strength of the breaking of chirality. 
It is important to stress that this is 
not true if there are gauge anomalies. If the theory is anomalous fermion 
loops induce gauge-breaking effects of $O(1)$ that cannot be subtracted. 
No approximate gauge invariance is possible in this case. In section
2, we will review the details of our two-cutoff formulation.

{\it The Foester, Nielsen and Ninomiya (FNN) mechanism of dynamical restoration
at large distances of a lattice gauge symmetry that is mildly broken at 
short distances \cite{fnn}}. Even though the two-cutoff construction will have 
an approximate gauge invariance, in any real simulation the ratio $f/b$ will be
finite. In this situation there is always some residual breaking 
of gauge invariance and it is then essential that the lattice model at finite
$f/b$ be in the same universality class as the model with $f/b = 0$ \footnote{
This might require a non-trivial tuning 
of the ratio $f/b$ as the continuum limit $b/\xi\rightarrow 0$ and/or the 
infinite volume limit $b/L \rightarrow 0$ is approached.}.   
 In section 3, we will review the FNN mechanism and 
extend the reasoning to the particular case of chiral gauge theories. 

In section 4, we will present the first numerical test of the TC 
method in a simple $U(1)$ chiral model in 2D. 

\section{Two-Cutoff Lattice Formulation}

In \cite{us1} and \cite{us2}, we presented a two-cutoff lattice construction of 
chiral gauge theories. Two different proposals with a similar philosophy 
were given in \cite{frolov,thooft}. Another earlier idea, in which 
fermions regulated in the continuum are coupled to interpolated lattice
gauge fields \cite{cont}, might also lead naturally to a TC construction. 

The advantage of using two-cutoffs, 
as we have explained, is precisely 
to make the violations of gauge invariance small. With 
one-cutoff a non-perturbative tuning of counterterms  would
be needed to achieve this \cite{roma}. 
The way in which the cutoff separation is implemented is by the use of two
lattices. The gauge degrees of freedom are the Wilson link variables of 
a Euclidean lattice of spacing $b$, ${\cal L}_b$. We will call $s$ the
sites of the $b$-lattice. The gauge action is the standard one: 
\begin{eqnarray}
S_{g} = \frac{2}{g^2} \sum_s \sum_{\mu<\nu} [I - \frac{1}{2}(U_{\mu\nu}+ U_{\mu\nu}^\dagger)] 
\nonumber\\
U_{\mu\nu} \equiv U_\mu(s) 
U_\nu(s+\hat{\mu})
U_\mu^{-1}(s+\hat{\nu}) U_\nu^{-1}(s).
\label{gauact}
\end{eqnarray}
Fermions on the other hand 
live on the sites of a finer lattice ${\cal L}_f$ (some integer subdivision of ${\cal L}_b$). We refer to the $f$-lattice sites as $x$. 
In order to decouple the unavoidable doublers, a Wilson term is included in 
the fermionic action. For each charged chiral fermion, a singlet of the opposite chirality is needed. 
The fermions are coupled to the gauge fields through a standard
lattice gauge--fermion coupling on the $f$-lattice. The link variables on ${\cal L_f}$
are obtained through a careful interpolation of the 
real dynamical fields, i.e. $U_{\mu}(s)$. As long
as the interpolation is smooth, it is clear that the separation of scales is
achieved in this construction, since the high-momentum modes of the gauge 
fields
are cut off at the scale $b^{-1}$, while the fermions can have momenta of 
$O(f^{-1})$. 
 The lattice action for a charged left-handed fermion field is then given by
\begin{eqnarray}
{\cal L}_{fermion} & = & \frac{1}{2} \sum_{\mu} \bar{\Psi} \gamma_{\mu} 
[ ( D^+_{\mu} + D_{\mu}^-) P_L + (\partial^+_{\mu}+\partial^-_{\mu}) P_R] \Psi\nonumber\\
& + & y \;\bar{\Psi} \Psi - \frac{r}{2} \bar{\Psi} \sum^{d}_{\mu=1} \partial_{\mu}^+ \partial^-_{\mu} \Psi \equiv \bar{\Psi}\; {\hat D}\; \Psi,
\label{u1}
\end{eqnarray}
where the covariant and normal derivatives are given by $D_{\mu}^+ \Psi(x) = u_{\mu}(x) \Psi(x+\hat{\mu}) - \Psi(x)$, $D_{\mu}^- \Psi(x) = \Psi(x) - u^\dagger_{\mu}(x-\hat{\mu}) \Psi(x-\hat{\mu})$, $\partial_\mu^+ = D_{\mu}^+|_{u=1}$ and 
$\partial_\mu^- = D_{\mu}^-|_{u=1}$ and we have introduced a bare mass term
for later use. As explained above, the $u_{\mu}(x)$ link variables are interpolations of the real dynamical fields $U_\mu(s)$.
When we refer to $f$-lattice quantities we use $f=1$ units and when we
refer to $b$-lattice quantities we use $b=1$ units for notational simplicity.
This should create no confusion. 

The TC lattice path integral is defined to be
\begin{eqnarray}
Z = \int_{{\cal L}_b} {\cal D}U \;\;\; e^{-S_{g}[U]}\;\; e^{\Gamma[u[U]]} \;\;
e^{-\bar{\eta}\; G[u[U]]\; \eta},
\label{pi} 
\end{eqnarray}
where ${\cal D}U$ is the standard $b$-lattice gauge measure, $G[u[U]] \equiv 
\hat{D}^{-1}$, and we define $\Gamma[u[U]]$ in the following way:
\begin{eqnarray}
\Gamma[u[U]] \equiv Phase({\det}_f(\hat{D})) \;\; \sqrt{{\det}_f(\not\!\!D)},
\label{eff}
\end{eqnarray}
where $\not\!\!D$ is the standard 
Wilson-Dirac operator on the $f$-lattice (i.e. the naive Dirac operator
plus a covariant Wilson term). The reason 
for this choice as opposed to the obvious one, i.e. $\Gamma[u[U]] = \det_f(\hat{D})$, 
is that in this way there is no breaking of gauge invariance in the real part
of the effective action. This definition is justified from the equivalent formal relation between the corresponding continuum operators \cite{al}. 
If the
real part broke gauge invariance, then it would be necessary to subtract several gauge non-invariant local  
counterterms generated by fermions at one loop \cite{us1}. 
Even though, in the TC method, only a one-loop subtraction 
would be needed (as opposed to the non-perturbative tuning required in the Roma 
approach), it is nevertheless a more difficult procedure 
 \footnote{For the particular 
interpolation we constructed in \cite{us2}, there could be
difficulties with this subtraction related to the 
discontinuities of the interpolated field across the boundaries
between $b$-hypercubes. See \cite{us1}.}, and thus we prefer the choice 
(\ref{eff}), where
no subtraction is needed. 

It was shown in \cite{us1} that, at the one-loop
level, under an infinitesimal gauge transformation $\omega$ on the $f$-lattice,
\begin{eqnarray}
\delta_{\omega} \Gamma[u] = Consistent \; Anomaly + O(f^2). 
\end{eqnarray}
Thus if gauge anomalies cancel, as we will assume the case to be, the 
fermionic effective action is gauge-invariant up to irrelevant terms.
Now, the power of the TC method is that these terms, which in OC constructions
 would induce $O(1)$ effects at higher orders, in this case become at most 
$O(f/b)^2$ to all orders, because gauge boson integration cannot bring
back powers of $1/f$, but only $1/b$. In order to prove this 
statement, we have to be more explicit about the
interpolated field, $u_\mu(x)$, which is a complicated function of the 
$b$-lattice gauge fields $U_\mu(s)$.

For a detailed explanation of how to construct such an interpolation for
the general case of a non-Abelian gauge theory in 4D, the reader is refered
to \cite{us2}. I just list here the properties that this interpolation must
satisfy in order to achieve the cutoff separation. 
\begin{itemize}
\item The interpolation is of course 
gauge-covariant under gauge transformations on the $b$-lattice, i.e. there
exists a gauge transformation on the $f$-lattice, $\omega(x)$, such that
\begin{eqnarray}
u_{\mu}[U^\Omega](x) = \omega(x) \; u_{\mu}[U](x) \; \omega^\dagger(x+\hat{\mu}) \equiv u_{\mu}^\omega(x).
\label{gt}
\end{eqnarray}
In this way, any functional that is gauge-invariant on the $f$-lattice is 
automatically gauge-invariant under the $b$-lattice gauge transformations 
(notice that the relevant gauge symmetry is that on the $b$-lattice). 

\item The interpolation must be 
covariant under the remaining discrete $b$-lattice symmetries ($90^\circ$ 
 rotations, translations, spatial and temporal inversions). This is  
important to ensure that Lorentz symmetry is recovered in the continuum limit
$b\rightarrow 0$. 

\item The interpolation has to be smooth in order
to achieve the cutoff separation. In \cite{us1}, it was shown 
that a sufficient condition to ensure that the 
gauge symmetry violations induced by the Wilson term in (\ref{eff}) 
are suppressed by $O(f/b)$  
is that the interpolated field, in the limit $f\rightarrow 0$, describes a 
differentiable
continuum gauge field inside each $b$-lattice hypercube, and its transverse
components are continuous across hypercube boundaries. We refer to this 
property as transverse continuity. From it, one can easily deduce a bound
on the high Fourier modes of the interpolated field,
\begin{eqnarray}
|a_{\mu}(q)| \leq \frac{C}{|q_{\mu}|} \prod_{\alpha\neq \mu} \frac{1}{q_{\alpha}^2}  
\end{eqnarray} 
where $C$ is independent of $f$ and $u_{\mu} \equiv e^{i a_{\mu} f}$. 
Using this bound it can be  shown that
\begin{eqnarray}
\Gamma[u[U^\Omega]] - \Gamma[u[U]] = O(f/b)^2\nonumber\\ 
G[u[U^\Omega]] - G[u[U]] = O(f/b)^2, 
\label{appro}
\end{eqnarray}
where $\Omega$ is a $b$-lattice gauge transformation and, in the second equation,
we have assumed that the external fermion momentum is smaller than $O(1/b)$. The proof in \cite{us1} 
is just based on the derivation of bounds for the lattice integrals 
corresponding to all the terms in 
the perturbative expansions of $\Gamma[u[U^\Omega]]-\Gamma[u[U]]$ and $G[u[U^\Omega]]-G[u[U]]$. 

\item The gauge-invariant degrees of freedom of the interpolated field 
should be local functions of those of $U_{\mu}$, where by local we mean within distances of $O(b)$. The reason for this is clear. If the gauge-invariant
degrees of freedom of the interpolated
fields were not constructed locally from the $U{\mu}$, the interpolation 
could modify the non-local properties of the correlation functions 
of gauge-invariant quantities and thus the physics would be 
interpolation-dependent. What is not so obvious is whether the gauge-dependent
degrees of freedom of the interpolated field should also 
be constructed locally from the $U_{\mu}$.
The reason why this does not seem to be necessary is that, even if 
gauge-dependent
degrees of freedom of $u_{\mu}$ were correlated at large distances with respect 
to $O(b)$ owing to the interpolation, this would at most 
induce long-range correlations suppressed by $O(f/b)^2$ in the physical
sector, according to 
(\ref{appro}). We can then always choose the cutoff ratio to be as small as 
is necessary to ensure that these unphysical 
correlations are negligible (for instance at
any finite volume, we can choose the cutoff ratio small enough so that 
the long-range correlation,  induced by the 
non-locality of the interpolation, between two physical fields at 
points $x_i$ and $x_j$ is smaller than $\exp(- |x_i-x_j|/b)$). 
In general we expect that the less local the interpolation is, 
the smaller the cutoff ratio will have to be; so, for practical reasons, 
it would be much preferable that the interpolation be as local 
as possible.  Our interpolation in 
\cite{us2} has the property that all gauge-invariant degrees of freedom 
of the interpolated field (for instance all the Wilson loops) 
depend on the $U_{\mu}$ strictly locally. However,  
this is not the case for the gauge-dependent 
degrees of freedom in $u_{\mu}$ \cite{singular}. In general it
is expected that for any group $G$ in $d$ space-time dimensions, whenever any of the $\Pi_n(G)$  for $n \leq d-1$ is non-zero, the gauge-dependent 
degrees of freedom of the interpolated field is either singular or depends non-locally on the $U_{\mu}$ fields \cite{us2}. Our numerical results for $U(1)$ 
in 2D show, however, that 
the non-locality of the gauge-dependent degrees of freedom of $u_{\mu}$ 
is mild,  in the sense that the correlation length of the $u_{\mu}$ fields, 
averaged over random  $U_{\mu}$ configurations, is finite and of $O(b)$. 
However, we do not have a proof of this in the general case.

\end{itemize}

\section{Effective Gauge Invariance}

The TC formulation only ensures that there is an approximate gauge
invariance for small $f/b$, but in any simulation  this ratio is necessarily
finite, so our lattice action is not gauge-invariant. There are, however, good 
arguments to believe that a pure gauge theory with a mild breaking of gauge 
invariance
at short distances flows in the IR to a theory with an effective 
gauge symmetry and no extra light degrees of freedom. 
The argument of ref. \cite{fnn} starts by showing that any lattice 
gauge theory (for a compact group) that contains some terms that break the lattice gauge invariance is equivalent to a theory with an exact gauge invariance
and additional scalar degrees of freedom. This is simple to see. Let  us consider a lattice action that contains gauge-breaking interactions:
\begin{eqnarray}
S[ U ] = S_{g.i.}[U] + \delta S_{n.g.i.}[U].
\end{eqnarray}
The path integral is
\begin{eqnarray}
Z = \int {\cal D}U_{\mu} \;\; e^{-S_{g.i}[U]-\delta S_{n.g.i.}[U]}.
\label{ni}
\end{eqnarray}
Since the group is compact we can multiply $Z$ by the volume 
of the group $\int {\cal D}\Omega$, which is an irrelevant constant factor:
\begin{eqnarray}
Z = \int {\cal D}\Omega \int {\cal D} U_{\mu}\;\; e^{S_{g.i}[U]+\delta S_{n.g.i.}[U]}.
\end{eqnarray}
Performing a change of variables from $U_{\mu}$ to $U^{\Omega}_{\mu} \equiv 
\Omega(x) U_{\mu}(x) \Omega^\dagger(x+\hat{\mu})$ 
(i.e. the gauge-transformed variables under 
the lattice gauge transformation $\Omega$), and using the invariance of the
measure and $S_{g.i.}$ under a gauge transformation\footnote{We are on a lattice, so even if the fermions are chiral the fermion lattice measure is invariant
under the unitary transformation $\Omega$.}, we get
\begin{eqnarray}
Z = \int {\cal D}\Omega \int {\cal D}U_{\mu} \;\;e^{S_{g.i}[U]+\delta S_{n.g.i.}[U^\Omega]}.
\label{gi}
\end{eqnarray}
It is easy to see that (\ref{gi}) is gauge-invariant under a new gauge 
symmetry under which the field $\Omega$ transforms as
\begin{eqnarray}
U_{\mu}(x) &\rightarrow & \Phi(x) U_{\mu}(x) \Phi^\dagger(x+\hat{\mu}) \;\;\;\; \Omega(x)  \rightarrow  \Omega(x) \Phi(x)^\dagger.
\label{gtg}
\end{eqnarray}
This is because $U_{\mu}^\Omega$ is invariant under this 
transformation.
This theory is then a gauge theory with extra charged scalars in a non-linear
realization ($\Omega$ is unitary). These scalars
are the pure gauge transformations, which couple through the non-invariant terms in the original action. 
It is clear that in any successful proposal for regulating chiral gauge 
theories, these scalars must decouple from the light physical spectrum. In the 
case of a spontaneously broken gauge theory 
these degrees of freedom remain, since they become the longitudinal 
gauge bosons. 

The FNN conjecture is that (\ref{gi}) flows in the IR to the same point as 
the theory (\ref{ni}) with $\delta S_{n.g.i.} = 0$, provided that the 
strength of the interactions in $\delta S_{n.g.i.}$ is ``small''. 
Their argument is quite simple. Let us suppose that the characteristic coupling
of the  non-invariant terms is arbitrarily small at the cutoff scale. 
Then in the gauge-invariant picture 
 of the theory (\ref{gi}), this implies that the $\Omega$ fields are 
very weakly coupled. Consequently, the free energy will be maximized when the 
$\Omega$ variables are decorrelated at distances of the order of the cutoff (i.e. the lattice spacing)  or, 
in other words, these degrees of freedom are effectively very massive. 
Then it makes sense to integrate them out in order to obtain an effective 
theory at low energies. At distances larger than the lattice spacing, but 
still smaller than 
the correlation length of the gauge-invariant degrees of freedom, it can 
be argued that 
the effective action should be local, because the $\Omega$ integration does
not generate long-range correlations, 
 and exactly gauge-invariant. (It is clear
from eq. (\ref{gi}) that if we perform the integration over $\Omega$, what 
remains is an exactly gauge-invariant theory due to the exact symmetry
 (\ref{gtg}).) In this situation, the only effect of the $\Omega$ fields
is a renormalization of the gauge-invariant couplings in $S_{g.i.}$. 

In the case we are interested in, there are also fermions. 
The previous reasoning would imply that the IR limit of 
the model (\ref{pi}) would be equivalent to that of 
the original theory without a Wilson term (since this is the only term in 
$\delta S_{n.g.i.}$),
and thus with doubling \cite{testa}. The reason why do not expect this to 
be the case
in the TC formulation is that the $\Omega$ integration gives an action 
that is expected to be local at distances of $O(b)$, but not
at distances of $O(f)$. This implies that 
after integrating the $\Omega$ fields, the effective 
fermion action on the $f$-lattice is 
not local. This is presumably how it can evade NN. 
On the other hand, if the $\Omega$ fields decouple at distances
of $O(b)$, there must exist an effective action at the $b$-scale that 
involves only the light fermionic degrees of freedom and would be local 
and gauge-invariant. Of course, it should violate in some other way 
the conditions of the NN theorem. In particular, there is no reason to 
believe that it would be bilinear in the fermion fields. In this case,  
it would not be very useful in real simulations.

In any case, it is clear that in the presence of fermions, the decoupling
of the $\Omega$ fields is not enough to ensure the right particle content.
One should make sure that the light fermion spectrum after the $\Omega$ 
integration is the correct one. 
It is easy to show that also our TC formulation of (\ref{pi}) is 
equivalent to a new theory with additional scalar degrees of freedom and an
 exact gauge symmetry. We will call this reformulation the Wilson-Yukawa 
picture, for reasons that will become clear. 
Following the previous steps  and using the property (\ref{gt}), 
the model (\ref{pi}) is exactly equivalent to
\begin{eqnarray}
Z = \int_{{\cal L}_b} {\cal D}\Omega \; \int_{{\cal L}_b} {\cal D}U \;\;\; e^{-S_{g}[U]} \;\;e^{\Gamma[u^\omega[U]]} \;\; e^{-\bar{\eta} G[u^\omega[U]] \eta}, 
\label{twocutwy}
\end{eqnarray}
where the $\omega$ fields are defined by (\ref{gt}) and are coupled uniquely 
through the Wilson and mass terms. The $\omega$ fields are functions of
the  $\Omega$ and $U_{\mu}$ fields and the interpolation procedure, and can be 
interpreted as smooth interpolations of the $\Omega$ fields to the $f$-lattice.
The transverse continuity property of the interpolation also implies that the
$f\rightarrow 0$ limit of $\omega$ is a differentiable field inside
each $b$-hypercube and continuous across $b$-boundaries \cite{us2}. Its
high Fourier modes $q > b^{-1}$ are then strongly suppressed,
\begin{eqnarray}
|\omega(q)| \leq  \prod_{\mu} \frac{C'}{q_{\mu}^2}.
\label{omq}  
\end{eqnarray}
Except for our particular choice (\ref{eff}) of the real part of the fermionic 
effective action, the model 
(\ref{twocutwy}) is a TC formulation of the Wilson-Yukawa models 
studied in \cite{wy}. In the quenched approximation, they only differ in 
the fact that the scalars in (\ref{twocutwy}) have a smaller momentum cutoff than
the fermions, but as we will see this turns out to be an essential difference. 

It is easy to see that (\ref{twocutwy}) has an exact gauge symmetry:
\begin{eqnarray}
U_{\mu}(x) &\rightarrow & \Phi_L(x) U_{\mu}(x) \Phi_L^\dagger(x+\hat{\mu}) \;\;\;\;\Omega(x)  \rightarrow  \Omega(x) \Phi_L(x)^\dagger, \nonumber\\
\Psi_L &\rightarrow & \phi_L(x) \Psi_L(x), \;\;\;\;\;\;\;\; \Psi_R  \rightarrow   
\Psi_R,
\label{tcgt}
\end{eqnarray}
where $\phi_L$ is a functional of $\Phi_L$ and $U_{\mu}$ defined by the 
condition,
\begin{eqnarray}
u_{\mu}[U^{\Phi_L}](x) = \phi_L(x) \;u_{\mu}[U](x) \;\phi_L^\dagger(x+\hat{\mu}).
\end{eqnarray}
There is also a global symmetry, under which
\begin{eqnarray}
\omega & \rightarrow & \phi_R\; \omega\; \nonumber\\
\Psi_R & \rightarrow & \phi_R\; \Psi_R.
\end{eqnarray}
The new gauge invariance comes at the expense of having unphysical 
degrees of freedom, $\omega$. As we have explained, in order to get
a chiral gauge theory, the $\omega$ fields must
decouple. This is expected to be the case in the TC formulation, because, thanks to the property (\ref{appro}), the FNN conditions are satisfied, i.e. 
the $\omega$ fields are weakly coupled to the gauge degrees of freedom and
to all low-momentum fermion modes (these interactions are suppressed by $O(f/b)^2$, except the coupling to the fermions with large momenta).
 For small enough $f/b$, the $\omega$ fields
are thus expected to decouple 
from the light spectrum. It is however possible that  
$f/b$ should scale with $\xi/b$ (correlation length of the gauge-invariant
degrees of freedom in lattice units) and
$L/b$ to ensure the decoupling of the $\omega$ fields at
large distances as the 
continuum limit is approached. Unfortunately to settle this question numerically we would need to be able to deal with complex actions, and this is 
still an open technical problem.  Nevertheless, for every finite $L/b$ and 
$\xi/b$, there must exist a small enough $f/b$, which ensures the decoupling of the $\omega$ fields, so that the low-energy theory has an 
effective gauge invariance and no extra scalar degrees of freedom. 

We also should make sure that the light-fermion spectrum is the 
correct one. For this it is necessary that the doubler modes be heavy
and that the right-handed fields in ${\hat D}$ and $\not\!\!D$ decouple. 
Let us first consider $\not\!\!D$.  To all orders in perturbation 
theory and in the TC case, 
the finite loops contributing to $\log(\det_f({\not\!\!D}))$, in which
a doubler mode propagates are suppressed at least by $O(f/b)^2$, because
the doubler mass if of the order of $1/f$. If the bare mass vanishes, also 
those contributions in which both the left-handed light mode and the 
right-handed one propagate are also suppressed by $O(f/b)^2$ terms, because
they necessarily involve a Wilson coupling. 
The only unsuppressed contributions are
those in which only either the left-handed light mode or the right-handed one propagate. The real part of these two contributions to 
 $\log[\det_f({\not\!\!D})]$ are equal, so the 
square root in (\ref{pi}) ensures that only the 
contribution from the physical field is taken into account. Notice that
this would not be true in a OC construction, in which there are non-local
and unsuppressed interactions  coming from fermion loops in which both 
L and R light modes propagate (presumably a tuning of the bare mass would be 
needed in that case to ensure the L/R decoupling). 

Secondly we should consider the chiral operator
${\hat D}$. The dependence on $\omega$ makes the analysis more subtle in this case. In the Wilson-Yukawa picture, the chiral operator 
corresponds to the action
\begin{eqnarray}
{\cal L}^{wy}_{fermion}  =  \frac{1}{2} \sum_{\mu} \bar{\Psi} \gamma_{\mu} 
[ ( D^+_{\mu} + D_{\mu}^-) P_L + (\partial^+_{\mu}+\partial^-_{\mu}) P_R] \Psi\nonumber\\
 +  y \bar{\Psi} (\omega^\dagger P_R + \omega P_L) \Psi - \frac{r}{2} \left ( \bar{\Psi} \omega^\dagger P_R \sum^{d}_{\mu=1} \partial_{\mu}^+ \partial^-_{\mu} \Psi + \bar{\Psi}P_L \sum^{d}_{\mu=1} \partial_{\mu}^+ \partial^-_{\mu}  \omega \Psi \right ).
\label{u2}
\end{eqnarray}
For $r=1$, the $\omega$ fields are strongly coupled to the 
fermions with large momenta, which implies that even if the 
$\omega$ fields actually 
decouple at large distances, we cannot simply read the light-fermion
spectrum from the Lagrangian (\ref{u2}). 
In general, we do not know what the light-fermion spectrum is 
(after all we have never
solved a chiral gauge theory non-perturbatively); however, it is
clear that in the limit in which we set the gauge coupling to zero, $u_{\mu} = 1$ in (\ref{u2}), the 
light-fermion spectrum should contain two (for each charged fermion in the theory) free, undoubled massless fermions, 
with chiral quantum numbers under the residual global symmetry,
\begin{eqnarray}
\omega & \rightarrow & \phi_R\; \omega\; \phi_L^\dagger \nonumber\\
\Psi_L &\rightarrow & \phi_L\; \Psi_L \nonumber\\ 
\Psi_R & \rightarrow & \phi_R\; \Psi_R.
\label{chi}
\end{eqnarray}
This expectation was not realized in any region of the Yukawa-coupling phase 
space $(y, r)$ in OC formulations, even in the quenched approximation 
\cite{wy}. The generic phase diagram found in those studies is depicted
in Fig. 1.
\begin{figure}
\begin{center}
\mbox{\epsfig{file=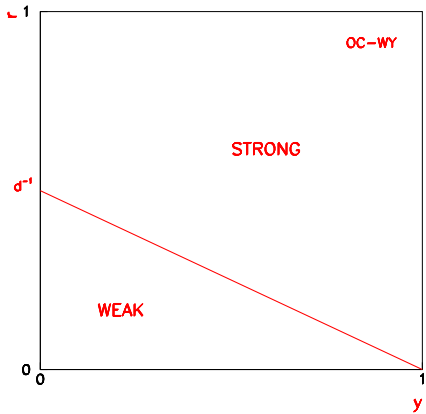,width=2.5in,height=2.5in}}
\end{center}
\caption[]{Sketch of the Yukawa phase diagram for a OC Wilson-Yukawa model.}
\label{fig1}
\end{figure}
 Two phases were found. In the weak phase $y + d\; r \leq 1$, the fermion masses behave as in perturbation theory in the continuum, i.e.
\begin{eqnarray}
m \sim \langle \omega \rangle,
\label{we}
\end{eqnarray} 
both for doublers and light modes. If the $\omega$ fields have only 
short-ranged correlations \footnote{In fact, OC Wilson-Yukawa models contained
a kinetic term for the scalar fields, $\kappa \sum_{x,\mu}\;( \omega^\dagger(x) \omega(x+\mu) + h.c.)$. In this case the correlation length is controlled 
by $\kappa$. For $\kappa < \kappa_c$, the $\omega$ correlation length is
finite and $\langle \omega \rangle = 0$.}, $\langle \omega \rangle = 0$, and the light-fermion spectrum is composed of massless fermions, but there is doubling. 
However, also a strong phase was found for $y + d\; r \geq 1$ where 
fermions get masses
proportional to a chirally invariant condensate:
\begin{eqnarray} 
m \sim \langle {\mbox Re}[ \omega(x) \omega^\dagger(x+\hat{\mu})\;] \rangle^{-1/2}.
\label{stro}
\end{eqnarray}
This condensate is non-zero in general,  and thus all fermions get masses
of the order of the cutoff in this phase, even though chiral symmetry is 
not broken. Since the couplings to the 
condensate are different for the light and doubler modes, there is a
splitting in the masses of the two sectors, and in particular the light-fermion masses can be tuned to zero (at least some of them) by a tuning
of the bare mass $y$ to some critical value, while the doublers, whose masses
are proportional to $r$, remain massive. The problem in this phase
is that the light-fermion spectrum is vector-like. The reason for this is 
that, owing to the strong Wilson-Yukawa couplings, mirror states can be formed, 
which are composites of the original fermions and the scalars. In particular, 
OC studies showed strong evidence that the composite Dirac field
\begin{eqnarray}
\Psi^{(n)}\equiv \omega \Psi_L + \Psi_R
\label{neu}
\end{eqnarray}
actually formed. Since it transforms vectorially under the global 
group, a Dirac mass is not incompatible with the exact chiral symmetry
(\ref{chi}). This field is called neutral, because it is neutral 
under the $U(1)_L$, which is the group that is gauged when $g > 0$.
There is also a charged Dirac fermion, 
\begin{eqnarray}
\Psi^{(c)}\equiv \Psi_L + \omega^\dagger \Psi_R,
\label{cha}
\end{eqnarray}
which is neutral under the $U(1)_R$ and transforms vectorially under
the $U(1)_L$. OC studies showed evidence \cite{gpr} that the charged composite
fermion 
is not formed and the state that propagates in this channel is a two-particle state, $\omega^\dagger \Psi^{(n)}$. In any case, the light-fermion spectrum is not the expected one. Either there are no light 
charged fermions or they couple vectorially to the gauge field. These models 
do not give rise to a chiral gauge theory. 

Again, the TC formulation changes this picture 
considerably. The main difference is that the high Fourier modes
of the $\omega$ field are suppressed (\ref{omq}) which implies that
the coupling of the scalars to the light-fermion mode $p\rightarrow 0$ 
is weak for $r=O(1)$ and $y \ll
 1$. 
\begin{figure}
\begin{center}
\mbox{\epsfig{file=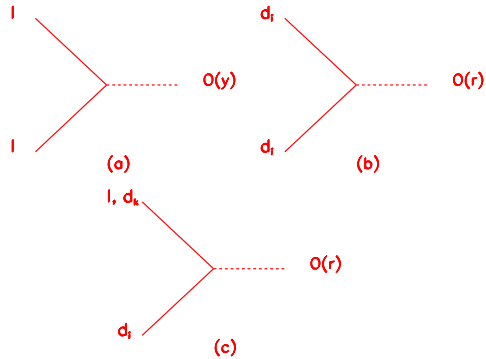,width=2.8in,height=2.8in}}
\end{center}
\caption[]{Strength of the couplings of the light $l$ and doubler modes $d_{i}$ to the
scalar field $\omega$ (dashed line) in the OC case.}
\label{fig2}
\end{figure}
Then, the light 
lattice modes of the original fermion fields, $\Psi_L$ and $\Psi_R$, 
are expected to remain massless according to (\ref{we});
in other words the poles at 
$p\rightarrow 0$ of the corresponding propagators are not lifted by 
the interaction with the $\omega$ fields, because this interaction is weak. 
The reason why this reasoning fails in OC formulations 
is that the light-fermion modes have strong couplings to the scalar
and other doubler
modes ((c) in Fig. 2). These couplings cannot be treated perturbatively. 
In the TC 
formulation, these mixed couplings are small, because they involve
high Fourier modes of the scalar field, which are strongly suppressed thanks
to the cutoff separation. On the other hand, the fermion doubler poles 
are still
strongly coupled to $\omega$ (the diagonal couplings, (b) in Fig. 2, are the same
for OC and TC) for $r = O(1)$ and we 
expect them to get masses according to (\ref{stro}). Thus the fermion
propagators will not have doubler poles. 

\begin{figure}
\begin{center}
\mbox{\epsfig{file=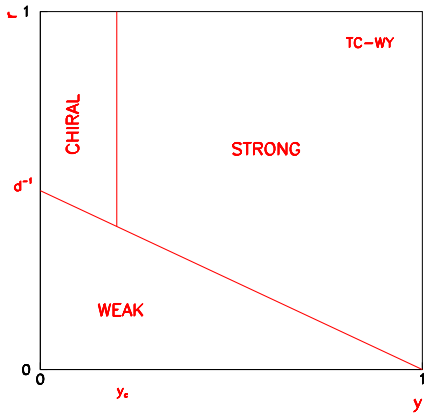,width=2.5in,height=2.5in}}
\end{center}
\caption[]{Sketch of the expected 
phase diagram for a TC Wilson-Yukawa model. The line separating the strong and chiral phases might not be vertical.}
\end{figure}
In summary, the expected 
Yukawa phase diagram for the TC formulation is depicted in Fig. 3. A truly 
chiral phase is expected for $r \geq 1$ and $y \ll 1$, in which the 
only massless modes are those corresponding to the poles at $p = 0$ of the 
fields $\Psi_L$ and $\Psi_R$, in (\ref{u2}). This is the correct fermion 
spectrum to define a chiral gauge theory. 

\section{A Two-Dimensional Model}

As we argued in the previous section, the first non-trivial test
of any lattice regularization of chiral gauge theories is to obtain,
at zero gauge coupling, undoubled chiral fermions at finite lattice
spacing. Extensive numerical studies in OC formulations \cite{wy} have
shown that this is highly non-trivial, already in the quenched approximation. 
In this section, I will review 
our numerical results \cite{hb} on the fermion spectrum in a $U(1)$ chiral 
model in 2D with the TC formulation, which support the expectation that a 
chiral undoubled spectrum is present, as shown in Fig. 3. 

At zero coupling, ${\not\!\!D}$ is trivial, since it reduces to 
the free operator. However, as 
we have seen, ${\hat D}$ is not, because it is coupled to the pure gauge
transformations, which are not suppressed at $g = 0$. $\hat D$ is obtained
from the action (\ref{u1}). The expression of  $u_{\mu}$ as a function
of $U_{\mu}$ can be found in ref. \cite{hb}.
Alternatively, in the Wilson-Yukawa picture, the $g=0$ limit 
corresponds to setting $U_{\mu}=1\rightarrow u_{\mu} =1$ in  (\ref{u2}) and 
what remains is a model of fermions coupled to the scalars $\omega(x)$ with 
a global $U(1)_L \times U(1)_R$ symmetry. 
(In the case $b=f$, this model is identical to the Wilson-Yukawa model 
considered in \cite{smit1} for $\kappa = 0$).
In this limit, the $\omega$ fields depend only on the $\Omega$. 
The reader is referred to \cite{hb} for the explicit formulae.

There are two non-trivial checks to be performed in this global limit. One concerns
the phase of the determinant of ${\hat D}[u^\omega]$, which should vanish in this limit, up to corrections of $O(f/b)^2$ according to
(\ref{appro}). The second concerns the fermion propagator 
$G[u^\omega[U]]$, which should describe 
two free massless undoubled fermions (for 
each charged fermion in (\ref{u2})), with 
the same quantum numbers under the $U(1)_L \times U(1)_R$ global symmetry 
as $\Psi_L$ and $\Psi_R$ in (\ref{u2}). 

Concerning the phase of the determinant, a numerical test of the expected 
power counting (\ref{appro}) is shown in Fig. 4. The three histograms 
correspond to the phase 
of the fermion determinant ($Phase(\det_f({\hat D}))$ in the background of 
$\omega$ configurations generated by interpolation of $\Omega$ configurations generated with the measure ${\cal D}\Omega$ in 
(\ref{twocutwy}) (i.e. randomly), and for three
values of the cutoff ratio $f/b=1, 1/4, 1/8$ and a $b$-lattice size of $L/b=4$. We refer to the TC lattice sizes with the notation $(L/b)_{(b/f)}$.
The charged-fermion content is anomaly-free: 
four left-handed fields with gauge charge 
$q_L=1$ and one right-handed one with charge $q_R=2$.
Clearly the change in the phase gets smaller with the ratio $f/b$, as expected, 
and this implies that the $\omega$ fields are weakly coupled in $\Gamma[u^\omega[U]]$. As we have argued, this is important to ensure that they will decouple from the light spectrum so that an effective gauge-invariant theory with no extra scalar
degrees of freedom will be obtained at large distances. 
\begin{figure}
\begin{center}
\mbox{\epsfig{file=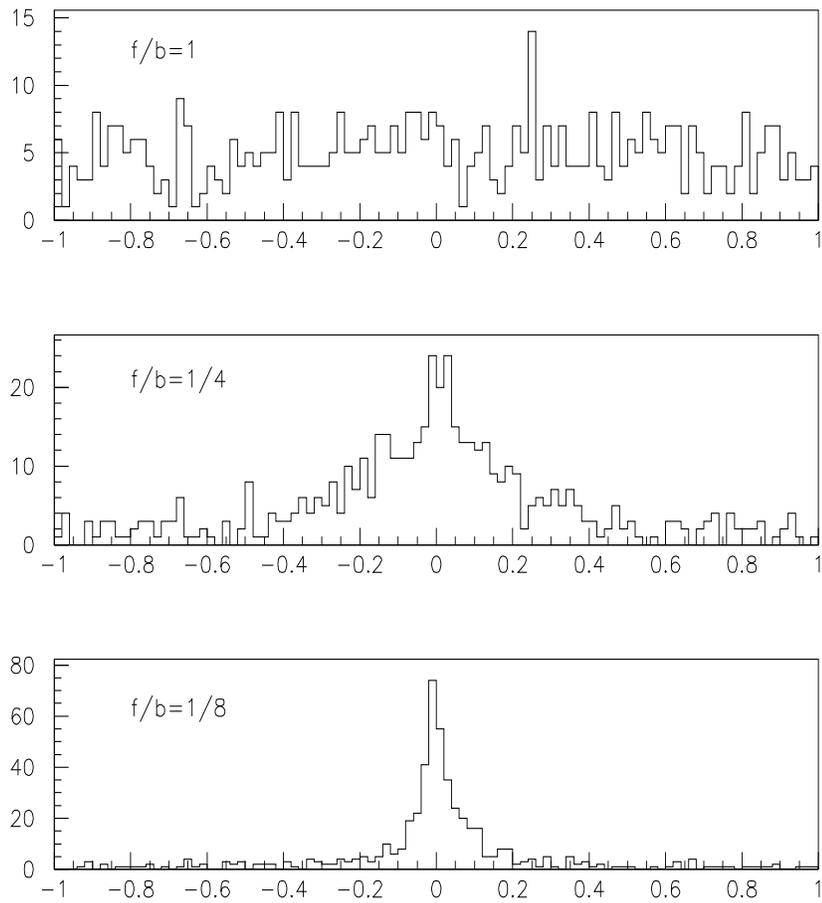,width=5in,height=6in}}
\end{center}
\caption[]{$Phase(\det_f({\hat D}))/\pi$ in the background of $\omega$ configurations obtained from the measure ${\cal D} \Omega$ in lattice sizes $4_1$, $4_4$ and $4_8$.}
\label{fig13}
\end{figure}

To study the fermion spectrum described by the propagator $G[u^\omega[U]]$, we are going to consider the quenched approximation. In this case, we can 
simplify the charged-fermion content
to just one left-handed fermion with charge 1, since the anomaly is not
present in this limit. The reason why this approximation is sensible is
because, as we have just seen, in the TC construction,
fermion-loop corrections to the $\omega$ effective Lagrangian are suppressed 
by powers of $f/b$, if gauge anomalies cancel. Thus, for 
small enough $f/b$, the scalar measure is going to be dominated by 
${\cal D}\Omega$ in the limit of zero gauge coupling.    

In the quenched approximation, the correlation length of the $\omega$ fields
only depends on the interpolation. In Fig. 5, we show the scalar 
propagator in time: 
\begin{eqnarray}
S_{\omega}(t) = \langle \omega^\dagger(x_1,x_2) \;\omega(x_1,y_2) \rangle  \;\;\; t=|y_2-x_2|,
\label{sca}
\end{eqnarray} 
where $x_1$ and $y_1$ are chosen randomly for every configuration. 
Even though the interpolation is not strictly local, the decay at large times 
of $S_{\omega}$ can be fitted to an exponential. The correlation 
length obtained is of $O(1)$ in $b$-units. 
This indicates that the scalar field decouples at large distances 
with respect to $b$ as a massive particle with a mass of the 
order of the boson cutoff. Provided $f/b$ is small enough, this will not
change when the quenched approximation is relaxed.
\begin{figure}
\begin{center}
\mbox{\epsfig{file=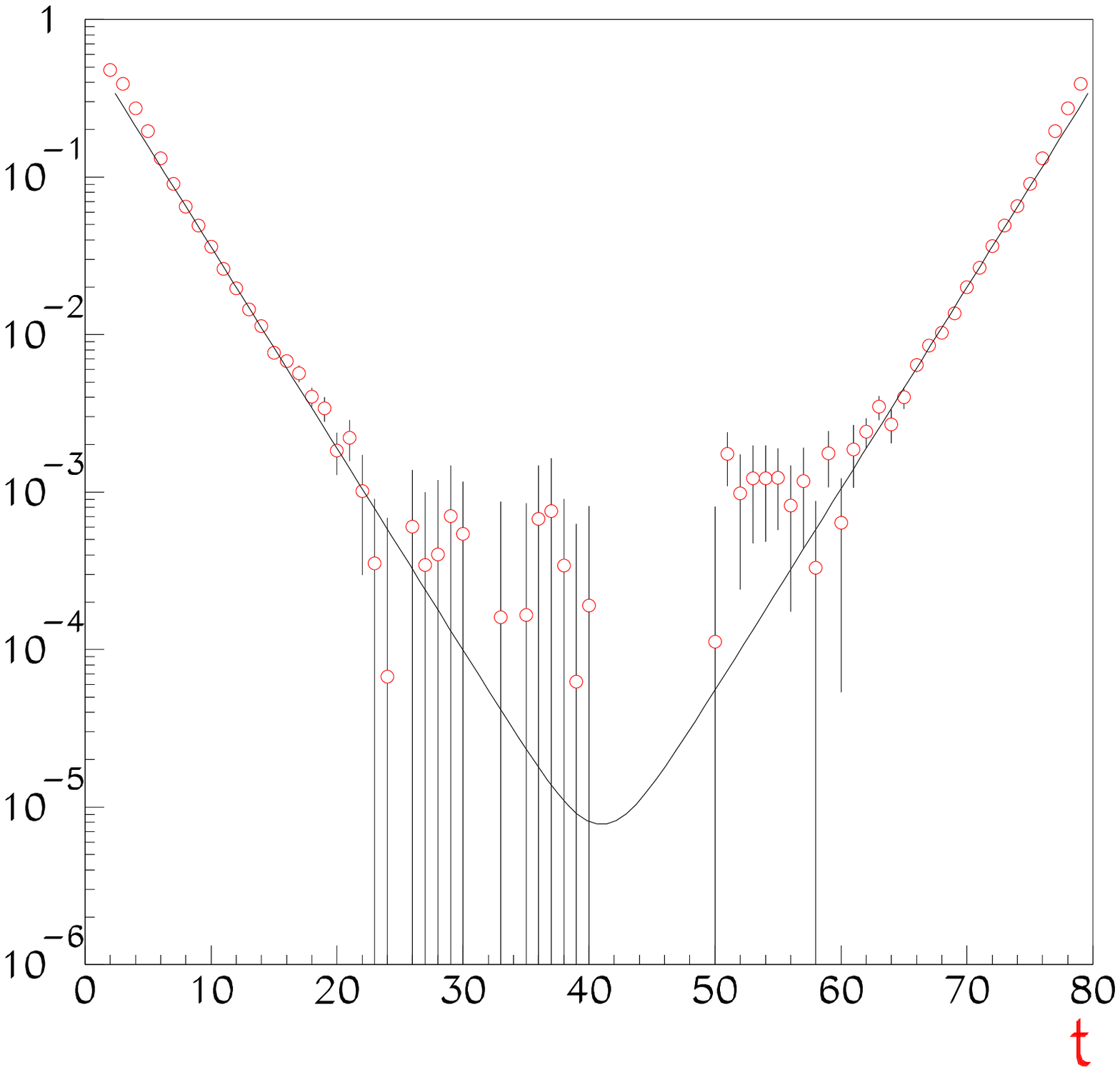,width=2.5in,height=2.5in}}
\caption[]{Scalar propagator (\ref{sca}) in a $20_4$ lattice.}
\end{center}
\end{figure}

\subsection{Fermion Spectrum}

We have computed the fermion propagators:
\begin{eqnarray}
S_{ij}^{a}(t) =  \langle \sum_{x_1,y_1} \;\; \Psi_i^{(a)}(x) \bar{\Psi}_j^{(a)}(y) \;\; e^{i p_1 (y_1 - x_1)} \;\; \rangle_{\Omega}  , \;\;\;\;\; t = |y_2 - x_2|,
\end{eqnarray}
where $t = 1,...,L$, the index $a$ refers to the neutral (n) (\ref{neu}), charged (c) (\ref{cha}) or 
physical ($\Psi_L+\Psi_R$) (p) fermion for $p_1=0$, and the corresponding spatial doublers (nd), (cd) and (pd) for
$p_1=\pi$. The indices $i, j = R, L$ to the different chiralities. The physical 
propagator corresponds to ${\hat D}^{-1}$, while 
the neutral and charged ones are
easily obtained from it. For every inversion, the time slice at the 
origin $x_2$ is chosen randomly. The number of sampled scalar 
configurations is typically of $O(2$--$5 \ 10^2)$. 
The matrix inversions have been performed with the conjugate-gradient 
method for several values of $y$ and fixed $r=1$. According to the expected
Yukawa phase diagram in Fig. 3, we should find a different spectrum for large
and small $y$.

At large $y \gg f/b$, we find that all the fermions are massive, as 
was also found in OC studies. Of course this phase has no physical interest, since 
all the particles, scalars and fermions, have masses of the order of the 
cutoff. However, it is useful 
to understand how fermion masses are generated even
if chiral symmetry is not broken. We find that the neutral propagator in 
momentum 
space is well reproduced by the hopping parameter expansion \cite{wy} (an expansion in $1/(y +2 r)$). To first order in this approximation, 
the neutral field is a free Wilson fermion with 
\begin{eqnarray}
y_{eff} = \frac{y}{z}, \;\;\;\;\;\; r_{eff} = \frac{r}{z},\nonumber
\end{eqnarray}
\begin{eqnarray}
Z_{L} = z^{-1},\;\;\;Z_{R} = 1,
\label{hopeff}
\label{effcou}
\end{eqnarray}
\begin{figure}
\begin{center}
\mbox{\epsfig{file=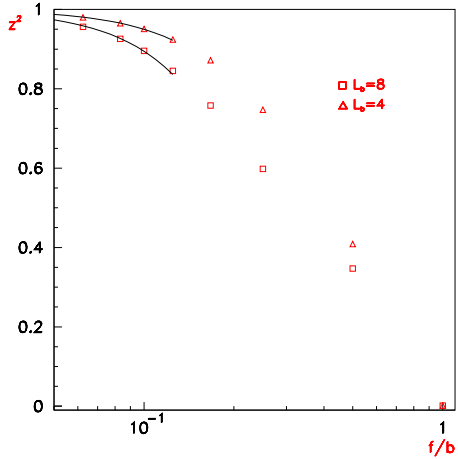,width=2.8in,height=2.8in}}
\end{center}
\caption[]{$z^2$ condensate for two lattice sizes $L/b=8, 4$ as a function of 
the ratio $f/b$. The lines are quadratic fits.}
\end{figure}
where $Z_{L,R}$ are the wave-function renormalization constants 
of the $\Psi^{(n)}_{L,R}$ fields, respectively, and $z$ is the strong 
condensate
\begin{eqnarray}
z^2 \equiv \;\langle {\mbox Re}( \omega(x) \omega^\dagger(x+\hat{\mu}) ) \rangle_{\Omega},
\end{eqnarray}
justifying formula (\ref{stro}). In Fig. 6, we show the value of 
this condensate for two $b$-lattice sizes as a function of $f/b$. It is 
easy to show that 
\begin{eqnarray}
\lim_{f/b\rightarrow 0} \;\; z^2 = \;\; 1.
\end{eqnarray}
In Fig. 7, we show two of the components of the neutral propagator $S^{(n)}_{LR}$ and $S^{(n)}_{RL}$ in momentum space, together with the fits to the 
free Wilson formulae for $y=1.5$ in an $8_8$ lattice. The fitted values 
of $y_{eff}$ and $r_{eff}$ are 
in good agreement with the hopping result (\ref{hopeff}).
\begin{figure}
\begin{center}
\mbox{\epsfig{file=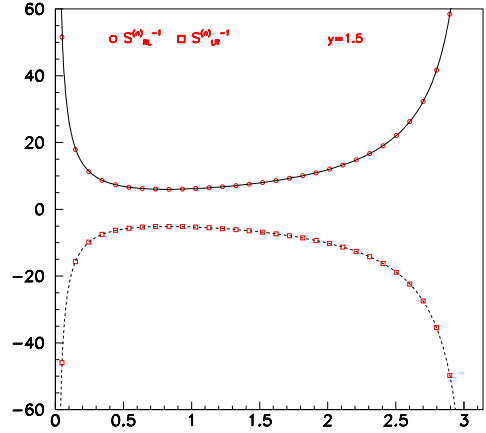,width=2.8in,height=2.8in}}
\end{center}
\caption{$(S^{(n)}_{RL,LR})^{-1}$ at $y=1.5$ and a lattice size of $8_8$.
 The lines correspond to the fits to the free Wilson fermion formulae. 
The parameters of the fits for $S^{(n)}_{LR}$ are
$Z_L^2 = 1.175(8)$, $y_{eff}=1.628(5)$ and $r_{eff}=1.083(4)$, while the hopping
result is $Z_L^2=1.177$, $y_{eff}=1.627$ and $r_{eff}=1.085$.}
\label{fig3}
\end{figure}
The neutral spatial doubler also shows good agreement with the hopping result.

The charged propagator (\ref{cha}) is also massive, as can be seen in Fig. 8. However, the
hopping expansion is not a good approximation in this case. Our data 
are also consistent with the charged channel being dominated by 
the two-particle state $\omega^\dagger \Psi^{(n)}$, as was found in OC 
studies \cite{gpr}.
\begin{figure}
\begin{center}
\mbox{\epsfig{file=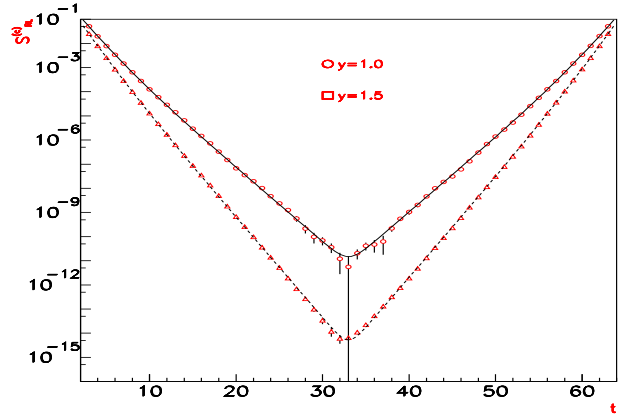,width=2.8in,height=2.8in,bbllx=60pt,bblly=-30pt,bburx=500pt,bbury=600pt}}
\end{center}
\caption[]{$S^{(c)}_{RL}$ propagators in time for $y = 1,1.5$. The lines
following the data points are the result of the two-exponential fits.}
\label{fig5}
\end{figure}

Finally, for the physical field, we have $S^{(p)}_{RL} = S^{(n)}_{RL}$ 
and $S^{(p)}_{LR}= S^{(c)}_{LR}$. The chirally breaking components
$S^{(p)}_{LL,RR}$ are compatible with zero, as they should be if chiral 
symmetry is exact. 

As we decrease $y \leq 1$, the different chiral components of both the 
light neutral and charged propagators start to differ. Two components get
 lighter than the others: $S^{(n)}_{RL} = S^{(p)}_{RL}$ and $S^{(c)}_{LR} = S^{(p)}_{LR}$ (i.e. 
the expected physical fields). 
There is a critical value of $y_c$ below which the two lighter
components $S^{(n)}_{RL}$ and $S^{(c)}_{LR}$ get massless at
finite lattice spacing. This is the onset of the announced chiral phase.  

\begin{figure}
\begin{center}
\mbox{\epsfig{file=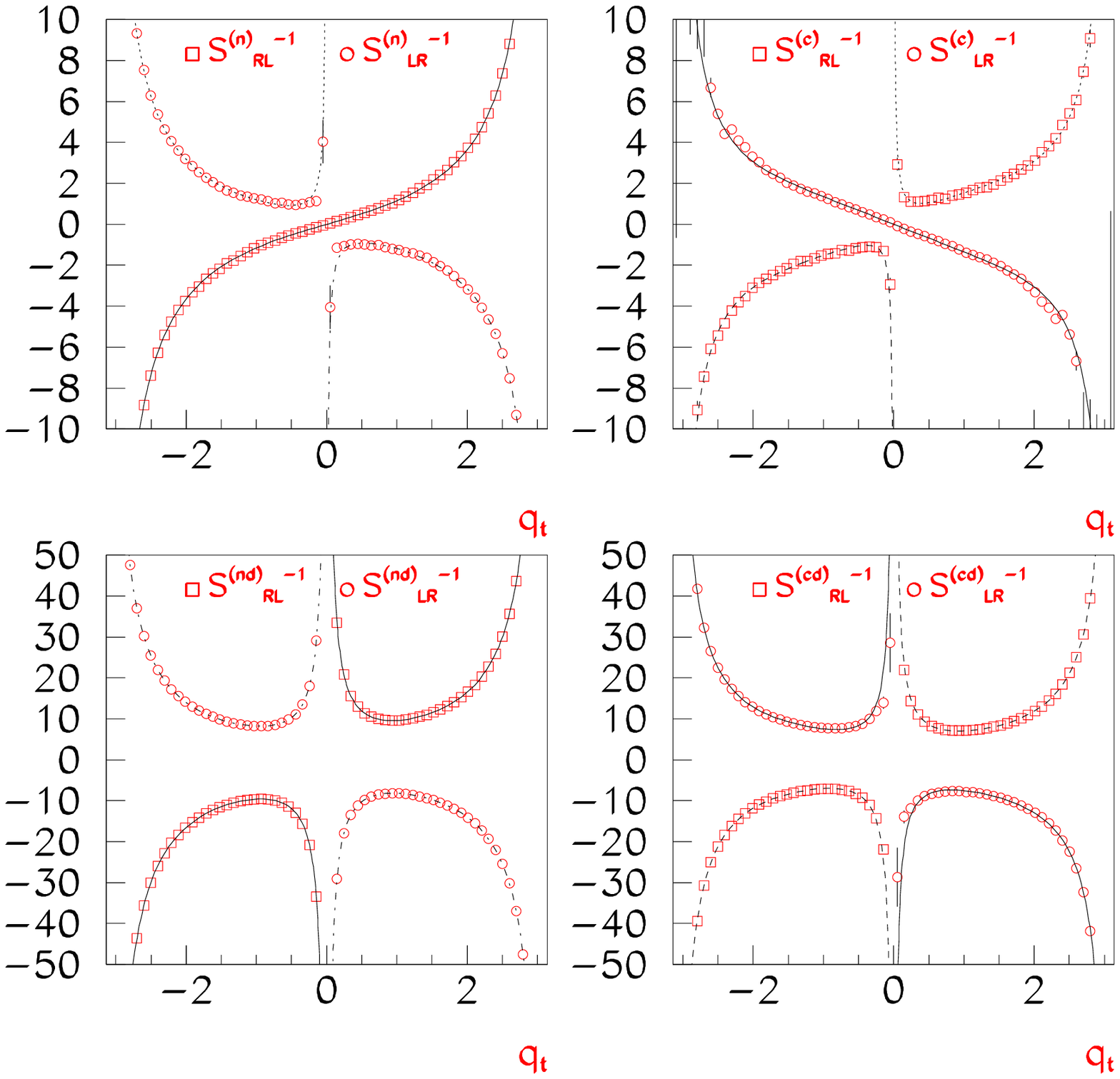,width=5in,height=5in}}
\end{center}
\caption[]{Inverse propagators $(S^{(n),(nd)})^{-1}$ and $(S^{(c),(cd)})^{-1}$ in momentum space in an 
$8_8$ lattice for $y=0.05$.}
\label{fig6}
\end{figure}
In Fig. 8, we show the $RL$ and $LR$ components of the neutral (n), charged
(c) and the corresponding spatial doublers (nd) and (cd) inverse propagators
in momentum space for $y < y_c$. Clearly there are only two components with poles at
$q=0$: $S^{(n)}_{RL} = S^{(p)}_{RL}$ and $S^{(c)}_{LR} = S^{(p)}_{LR}$, that
is the original fields in the Lagrangian (\ref{twocutwy}). These propagators 
show no other
poles in momentum space, as expected from undoubled fermions. None of 
the other components of the neutral and charged propagators, nor any of the 
doublers, have any pole in the whole Brillouin zone. So the only massless
modes are those expected.  The doubler propagators $\Psi^{(n)}$ and 
$\Psi^{(c)}$ behave as massive Dirac fields, supporting the expectation that
the doubler modes remain in the strong phase. 
The dependence on $y$ for $y < y_c$ is negligible. Also the volume 
effects  for fixed $f/b$ ratio are very small \cite{hb}. On the other
hand, the dependence on the ratio $f/b$ seems 
very important. We have found that, as this ratio 
increases, the splitting between the doubler and light sector decreases
dramatically. This indicates the importance of the cutoff separation to 
ensure the doubler-light splitting. 
\begin{figure}
\begin{center}
\mbox{\epsfig{file=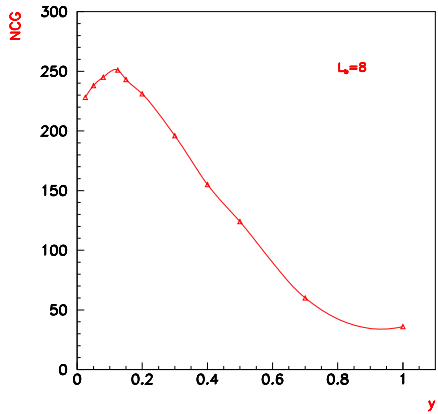,width=2.5in,height=2.5in}}
\end{center}
\caption[]{Number of conjugate-gradient iterations in the inversion of the 
neutral propagator as a function of $y$ in an $8_8$ lattice.}
\label{fig8}
\end{figure}

It is not clear what determines the value of $y_c$. 
There is one obvious candidate, which is the boson cutoff, i.e. $f/b$ in $f$-lattice units. We have monitored the number of conjugate gradient iterations
(NCG) required in the fermion matrix inversion, which is 
related to the minimum eigenvalue of the matrix. As $y$ decreases, the mass
of the fermions decreases, and so does the lowest eigenvalue of the fermion matrix, so that NCG 
grows. However, when we enter the chiral phase, the lowest eigenvalue is
 the IR cutoff since there are massless fermions, so the NCG should not grow
further. In Fig. 9 we show the NCG as a function of $y$. We find a maximum where
we expect $y_c$ to be, according to the behaviour of the propagators.
It is located around $f/b$ for this lattice size. 
 The physical picture behind this expectation is 
nothing but the FNN conjecture that at distances 
larger than $b/f$ in $f$-units, the scalars should decouple. The 
correlation length of the light fermions in the strong phase is $\sim y$, so 
when $y < f/b$ the scalars decouple from the light fermions and a chiral
phase appears.

More statistics will be needed to understand the properties of the transition
at $y_c$. In any case, the numerical evidence in this simple model 
supports the expectation that the chiral phase in Fig. 3 exists, which 
has the correct light fermionic degrees of freedom to obtain a chiral gauge theory. Similar conclusions are expected in 4D models.  

\section*{Acknowledgements}
I would like to thank my collaborators in the work reviewed here,   
Ph. Boucaud and R. Sundrum.

\end{document}